\documentclass[twocolumn,showpacs,pre,amsmath,amssymb]{revtex4}

\usepackage{graphicx}

\newcommand{\itd}[3]{\left(\partial #1/\partial #2\right)_{#3}}  
\newcommand{\infi}{^{_\infty}}

\begin{document}

\title{Non-Equilibrium Thermodynamic Description of the Coupling\\ 
 between Structural and Entropic Modes in Supercooled Liquids}

\author{
        R.~Di~Leonardo$^{1}$,
        A.~Taschin$^{2}$,
        M.~Sampoli$^{2,3}$,
        R.~Torre$^{2}$,
        G.~Ruocco$^{1}$,
        }

\affiliation{
    $^{1}$Dipartimento di Fisica and INFM, Universit\'a di Roma ``La Sapienza'', I-00185, Roma, Italy.\\
    $^{2}$Dipartimento di Fisica, LENS and INFM, Universit\'a di Firenze,  I-50019 Sesto, Firenze, Italy.\\
    $^{3}$Dipartimento di Energetica, Universit\'a di Firenze, via S. Marta, Firenze, Italy.
    }

\date{\today}

\begin{abstract}
The density response of supercooled glycerol to an impulsive
stimulated thermal grating ($q$=0.63 $\mu$m$^{-1}$) has been
studied in the temperature range ($T$=200$\div$340 K) where the
structure rearrangement ($\alpha$-relaxation) and
thermal diffusion occur on the same time scale. 
A strong interaction between the two modes occurs giving rise to  
a dip in the $T$-dependence of the apparent thermal
conductivity and a flattening of  the  apparent
$\alpha$-relaxation time  upon cooling. 
A non-equilibrium thermodynamic (NET) model for the long time response 
has been developed. 
The model is capable to reproduce the experimental 
data and to explain the observed phenomenology.

\end{abstract}


\maketitle

Whenever an inhomogeneous temperature or pressure field exists
inside a substance, heat and momentum will flow giving rise to
processes, like thermal diffusion and sound propagation, which drive
the system toward homogeneity. In a normal liquid, at low enough wave-vector q 
(i.e. at typical values of light scattering experiments), the
time-scales of the two processes are well separated, so that
sound propagation is adiabatic and thermal diffusion is isobaric.
Every other microscopic process evolves on such a fast
time-scale that it enters the dynamic equations simply by
determining the actual values of thermodynamic derivatives and
transport coefficients. 
The situation changes when a liquid is
supercooled below its melting temperature and the structural
relaxation time, $\tau_\alpha$, rapidly grows up upon
cooling. When $\tau_\alpha$ becomes of the order of magnitude of the sound wave
period, we observe phenomena as the sound velocity dispersion and
sound absorption. These phenomena have been widely investigated by
ultrasonic and Brillouin spectroscopies
and commonly described in terms of a relaxing bulk modulus or 
viscosity\cite{Mountain,zwanzig1}.
Upon further cooling, $\tau_\alpha$ reaches the time scale of the
thermal diffusion giving rise to a complex frequency dependent heat diffusion
that is observed experimentally by specific heat spectroscopy \cite{birge}
and forced Rayleigh scattering \cite{allain,koler}.
Though both viscosity and specific heat relaxations are
manifestation of the same microscopic process, there is no
commonly accepted formulation of the
dynamic equations in the region where structural relaxation and thermal
diffusion occur on the same time-scale.
The main difficulty arises when more than one single thermodynamic
derivative has to be generalized to have a frequency dependence.
To this respect NET provides a more fundamental
approach, compared to generalized hydrodynamics, since, once
the equation of state is properly written in a suitable extended parameter space,
the frequency dependence of generalized 
thermodynamic derivatives comes out naturally.
Moreover a thermodynamic approach could, hopefully,
be related to that thermodynamic picture of supercooled and glassy state 
which has recently been the subject of great theoretical and computational 
efforts \cite{sciorti}.
Unfortunately whether local thermodynamic equilibrium is still meaningfull 
in the supercooled regime and what are the crucial steps in extending the 
thermodynamic parameters space are still open questions.
 
In this work we present an Impulsive Stimulated Thermal
Scattering (ISTS) \cite{nelson_ISTS} study of liquid and
supercooled glycerol in a temperature range that covers the region
where the characteristic times of structural and entropic modes become 
similar.
A strong interaction between the two modes occurs and we have observed 
an apparent slow down of the steep increase 
of structural relaxation upon cooling, together with a marked non exponential decay
at long times where the ISTS signal is usually governed by thermal diffusion. 
The NET model we propose is based on local thermodynamic equilibrium 
in an extended parameter space\cite{degroot}.
Using literature data, we reproduced accurately the experimental ISTS responses at long times
and explained unexpected features such as the dip in the T-dependence 
of the thermal conductivity of various supercooled
liquids \cite{allain,koler,Torre}.

\begin{figure}
\centering
\includegraphics[width=.38\textwidth]{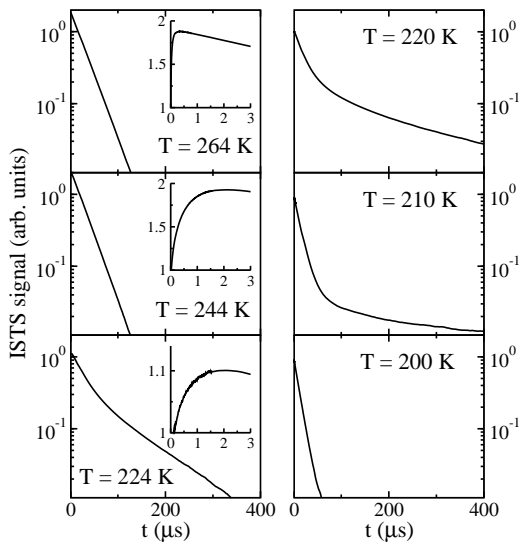}
\caption{\label{spectra}ISTS data of
supercooled glycerol at selected temperatures are reported in
logarithmic scale and normalized to unity at about $t=0.1\mu$s.
The inserts show a blow up of the short time region, where the
structural relaxation gives rise to a rising signal.}
\end{figure}

\begin{figure}[t]
\centering
\includegraphics[width=.38\textwidth]{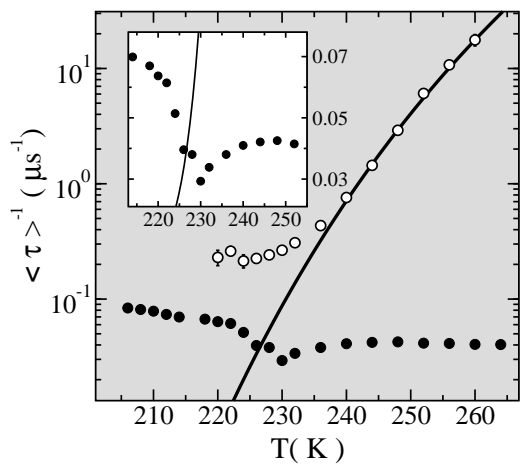}
\caption{\label{expgamma} Temperature dependence of charachteristic
rates for the short rising component -apparent structural relaxation- ($\circ$)
and  intermediate exponential decay -apparent thermal diffusion- ($\bullet$).
Solid line is the VTF law $\gamma_{_{VTF}}=\tau_0^{-1}
exp[- 2260 K /(T- 131 K)]$ }
\end{figure}

\begin{figure}
\hspace{-.7cm}
\includegraphics[width=.45\textwidth]{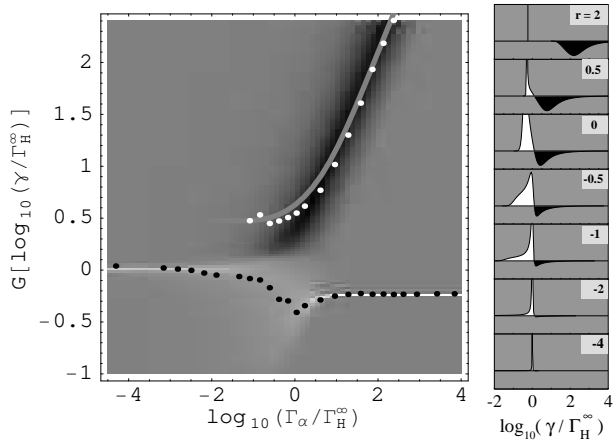}
\caption{\label{timedistr}Computed evolution of the distribution of rates $G$ as a function of the
logarithmic ratio $r=\log_{10}(\Gamma_\alpha/\Gamma^\infty_H)$ between structural 
relaxation characteristic rate and
the infinite frequency thermal decay rate. Circles in the left panel
represent the experimental values}
\end{figure}

Glycerol (99.5+\%,\;$<$0.1\% water, Fluka, glass transition temperature 185 K) was 
transferred under nitrogen into a Teflon-coated cell with movable windows
\cite{nelson_cella}. 
The cell was mounted on the cold finger of a
cryostat and outfitted with resistive heaters. A platinum resistance
thermometer was immersed in the sample and the temperature was kept
stable within $\pm$0.1 K.
In the present  ISTS experiment two infrared ($\lambda=1064$ nm)
short ($\sim 100$ ps) laser pulses
cross each other in the sample volume 
at an angle of $\sim 6^\circ$ ($q=0.63 \mu m^{-1}$) and their 
interference produces an impulsive, spatially modulated, heating.
The amplitude of the resulting density grating is  
probed by a third CW laser beam ($\lambda=532$ nm)
impinging on the induced grating at the
Bragg angle. The intensity of the diffracted beam as a function 
of time is stored by a digital oscilloscope 
and averaged over many ($\approx$ 5000) pulses. 
The measured signal can be fully ascribed to density dynamics 
excited through heating   
since electrostrictive and birefringence
effects are negligible in glycerol \cite{Paolucci,taschin,Glorieux}.
Further experimental details are reported in
Ref.~\cite{Torre}.
ISTS data were collected in the temperature range $T$=200$\div$340 K. 
Signals as long as 1.6 ms were recorded with a
time resolution of 1 ns.
In the present paper we focus on the long time
part of ISTS signal ($t > 0.1\mu$s) where the acoustic transient
is over and structural and entropy modes evolve isobarically.
Selected ISTS data are reported in
Fig.~\ref{spectra} showing two time regions: ``short times'' ($t < 10\mu$s)
in the inserts and longer times in tha main panels. It can be noted that
at short times the amplitude of the density
grating (ISTS signal) increases with a stretched exponential law,
due to structural relaxation.
On lowering the temperature, the characteristic time of this rising
component stops to grow while its intensity vanishes.
At longer times the heat diffuses and the density grating decays to zero.
This long time decay, which in absence of coupling is exponential  
(due to the diffusive character of heat equation) \cite{nelson_ISTS},
now splits into two components.
The faster component is nearly exponential and its time constant
goes through a maximum and gives rise to a dip in the apparent 
thermal diffusivity, as already reported for OTP in \cite{Torre}. 
The slower component shows a strong non-exponentiality.
It flattens out and its intensity disappears
as the temperature is lowered. 
To be more quantitative,
we  fitted the data with two
stretched exponentials for the "structural" rising component
and the long time tail, and a simple exponential for the 
intermediate time component (apparent thermal diffusion).
The decay rates \cite{note1} of the short and intermediate
components are reported in Fig.~\ref{expgamma}.
At high temperatures, the structural relaxation decay rate
($\circ$) is described by a Vogel-Tamman-Fulcher (VTF) law (full line).
The parameters $B$=2260 K, $T_{VTF}$=131 K of VTF law, 
$\gamma_{_{VTF}}=\tau_0^{-1} exp[-B/(T-T_{VTF})]$ are taken 
from dielectric spectroscopy \cite{lunke} while $\tau_O=1.4\cdot 10^{-15}$s
is scaled to fit the data. 
At temperatures lower than $\sim$240 K
the short time rising component ceases to represent the structural
relaxation and flattens around a value of 4 $\mu$s. At about the
same temperature, the apparent thermal
decay rate deviates from the expected smooth behavior (see
the insert of Fig.~\ref{expgamma}) and exhibits a
dip at $T$ $\approx$230 K. The overall scenario depicted in
Fig.~\ref{expgamma} suggests the existence of an interaction
between the structural and thermal relaxation dynamics that
leads to a complex relaxation time pattern. 
In the following we will introduce a NET model
that will allow us to compute the temperature evolution of
this relaxation times pattern and find a very good agreement
with the experimental observations.
NET \cite{degroot} provides a very powerful
framework to study irreversible processes
such as heat conduction, diffusion and viscous flow, from an unified point
of view. However, in this formalism, it is not straightforward to
consider the non-exponentiality observed in the
structural relaxation dynamics. 
On the contrary, the presence of a large number of internal 
relaxing variables can be easily taken into account provided 
that local thermodynamic equilibrium is valid in the
extended parameter space.
Therefore, following Allain et al.\cite{allain_th}, we choose
to represent the observed non-exponentiality as the result of the superposition of $N$
linearly relaxing variables \cite{note3}. 
In this hypothesis the Gibbs free energy law per unit mass reads:
\begin{equation}
dg= v dp -s dT -\mbox{$\sum_{i=1}^N$} A^i d\xi^i
\end{equation}
where $p$ is the pressure, $s$ the entropy, $A^i$ the affinity 
of the $i^{th}$ relaxation process and
$\xi^i$ the correspondent progress variable (or order parameter).
As noted above, we are interested in the time region where
the pressure became and stays uniform. In this time
region, the linearized NET equations\cite{degroot} 
written in terms of the q-components of the thermodynamic variables 
(e.g. $T$ stands for $T(t)=\int exp(i{\mathbf {qr}})T({\mathbf r,t})$),
are:
\begin{eqnarray}
\nonumber
p&=&0 \\
\label{startingeq}
T_0\rho_0 (\partial s/\partial t)&=&-q^2\lambda T \\
\nonumber \partial \xi^i/\partial t&=&-\beta^i A^i
\end{eqnarray}
where $T_0$ ($\rho_0$) is the
average temperature (density), $\lambda$ the thermal conductivity
and $\beta^i$  phenomenologic constants.
The first of (\ref{startingeq}) comes from the linearized combined mass
momentum conservation laws \cite{notaLongTimeApprox}.
The second and third equations 
of (\ref{startingeq})
represent the energy conservation law
and the phenomenological relations for the relaxation processes respectively.
In order to close the above set of equations, we use 
the local thermal equilibrium in the extended parameter space:
\begin{eqnarray}
\nonumber
\rho&=&\rho(p,T,\xi^1,...,\xi^N)\\
s&=&s(p,T,\xi^1,...,\xi^N)\\
\nonumber
A^i&=&A^i(p,T,\xi^i)
\end{eqnarray}
For simplicity, we assume that the thermodynamic affinity
$A^i$ does not depend on $\xi^j$ for $j\neq i$. 
Differentiating the above equations and
substituting in (\ref{startingeq}) we obtain:
\begin{eqnarray}
\label{feq1}
\rho&=&-\rho_0 \alpha\infi T+(\rho_0^2 c\infi_p/T_0)
\mbox{$\sum_{i=1}^N$} \Delta^i (\xi^i_p/\xi^i_T) \zeta^i
\\
\label{feq2}
\partial T/\partial t&=&-\Gamma^\infty_H T-
\mbox{$\sum_{i=1}^N$} \Gamma^i_R \Delta^i (T -\zeta^{i})
\\ \label{feq3}
\partial \zeta^i/\partial t &=&-\Gamma^i_R (\zeta^i -T)
\end{eqnarray}
where we have introduced the following symbols:
\begin{equation}
\begin{array}{rclcrcl}
\alpha\infi&=&-\rho^{-1}\itd{\rho}{T}{p \xi}&\;&
c\infi_p&=&T_0\itd{S}{T}{p \xi}\\
\Delta^i&=&T_0 A^i_\xi \xi^{i2}_T/c\infi_p &\;\;&
A^i_\xi&=&\itd{A^i}{\xi^i}{p T}\\  
\xi^i_p&=&\itd{\xi^i}{p}{A^i T}&\;&
\xi^i_T&=&\itd{\xi^i}{T}{A^ip}\\
\zeta^i&=&\xi^i/\xi^i_T&\;&
\Gamma^\infty_H&=&\lambda q^2/\rho_0 c\infi_p\\ 
\end{array}
\end{equation}
The ISTS density response is obtained by solving (\ref{feq2},\ref{feq3})
for the initial condition $T(0)\neq 0$, $\xi^i(0)$=$0$ and then substituting
in (\ref{feq1}). 
To reduce the number of parameters, we use the simplifying assumption 
that $\xi_p^i/\xi_T^i$ is independent of $i$ \cite{allain_th}.
In that case, $\rho(t)$ can be written as
\begin{equation}
\label{density} 
\rho(t)/\rho(0)=-T(t)/T(0)-\frac{\Delta \alpha}{\alpha\infi}\frac{c\infi_p}{\Delta c_p}
\mbox{$\sum_{i=1}^N$} \Delta^i \zeta^i(t)
\end{equation}
where $\Delta\alpha$ and $\Delta c_p$ are the jump from the relaxed 
to the unrelaxed (with respect to the $\zeta^i$)
values of the corresponding thermodynamic derivatives.  
%
It can be easily shown that in the limit $\Gamma^i_R\ll
\Gamma^\infty_H$ - i.e. at  low T or high $q$ values-
this model predicts a density-density correlation function which
at long times decays as: 
\begin{equation}\label{modelfi}
\phi_q(t)\propto \mbox{$\sum_{i=1}^N$} \Delta^i
\exp[-\Gamma^i_R t]
\end{equation}
\begin{figure}
\centering
\includegraphics[width=.4\textwidth]{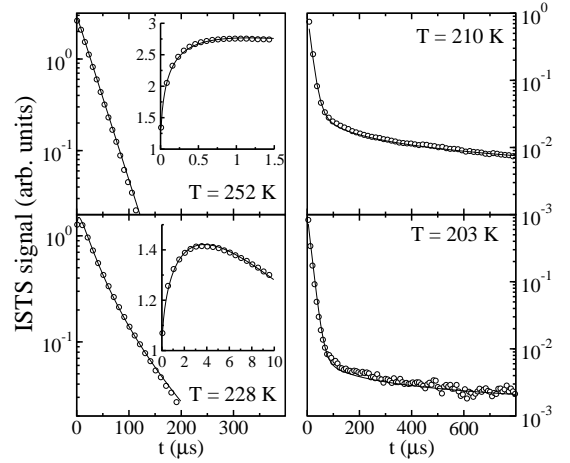}
\caption{\label{timecomp} ISTS data ($\circ$) and the
predicted signal (solid line) for four different temperatures.
The agreement is very good through the whole structural-entropic 
coupling region}
\end{figure}
We know from photocorrelation experiments, Mode Coupling Theory and
molecular dynamics simulations, that the above correlator is very
well described by a stretched exponential
$\exp[-(t/\tau_\alpha)^\beta]$ where $\beta$ is slightly changing
with temperature and $\tau_\alpha$ obeys a
VTF law. Relying on this considerations, 
we arbitrarily choose a distribution of $N$=150 \cite{noteN}
logarithmically spaced rates
$\Gamma^i_R$=$(\Gamma_\alpha) 10^{x_i}$, $x_i$=
-1+i/28, with weights $\Delta^i$ 
such that the sum
(\ref{modelfi}) reconstructs a stretched exponential with
$\beta$=0.65 \cite{birge,Paolucci}.
This determines the weights apart from a constant factor
which in turn can be easily fixed by the value of $\Delta c_p/c\infi_p=\sum\Delta^i$ 
form specific heat spectroscopy data\cite{birge}. 
The temperature simply changes
the value of $\Gamma_\alpha$ which 
is assumed to obey the already quoted VTF law.
We can assume
$\Gamma_H^\infty(T)$=$(c_p^0/c_p\infi) \Gamma_H^0(T)$ where
$\Gamma_H^0(T)$ is the extrapolation to the whole temperature
range of ISTS thermal decay rates at high temperature. Finally, from
$\rho(T)$ data across the glass transition temperature \cite{ubbe}
one finds $\Delta\alpha\sim 3.2 \alpha\infi$.
We are now left with no more free parameters: for each temperature we can
compute the  ISTS signal and compare it to
the experimental data. As examples, the results of this comparison
are shown in Fig.~\ref{timecomp} for four different temperatures,
showing an excellent good agreement between the data ($\circ$)
and the model (solid line) in the whole examined temperature
range. 
A complementary and more insightful way of representing complex time 
responses consists in performing an inverse Laplace transform analysis:
$I(t)=\int_{-\infty}^{\infty} G(\log\gamma) \exp[-\gamma t] \; d\log\gamma$.
%
%
In other words, one can think of the ISTS signal as a superposition of
exponentials and ask how the weight function $G$ evolves across
the coupling region.
In this representation it is easier to visualize
the analyzed phenomenology as the interaction between an exponential
thermal process and a broad distribution of relaxing variables. As a consequence,
the relation between the flattening in the "structural" 
rate, the dip in the thermal rate and the appearence of a long time tail becomes 
evident. 
Such an analysis has been carried out on the simulated 
signals and is summarized in Figs.~\ref{timedistr}.
The top figure in the right panel 
represents $G$ in the high temperature region where no
coupling is present. Structural relaxation manifests itself as a rising
(negative weight) stretched (broad distribution) exponential,
while at longer times (smaller rates) thermal diffusion
contributes to the signal with an exponentially (narrow) decaying
(positive weight) component. As the temperature is lowered, the broad
structural mode moves to shorter rates until the tail of its rate
distribution reaches the entropy mode time scale. As a result, the
time scale of the negative component ceases to vary 
and its intensity vanishes. On the other hand, the positive
component broadens moving to smaller rates and then splits into
two components: a narrow one which moves to larger rates, lowering the
temperature, and a broad one which becomes flatter and
flatter and decays to zero. 
In the left panel of Fig.~\ref{timedistr}, we report the
evolution of $G$ as a function of 
$log_{10}(\Gamma_\alpha/\Gamma^\infty_H)$. 
Black and white regions represent negative and positive weights respectively.  
The circles represent the normalized average rates for the
rising (white) and first part of decaying (black) portion of the experimental ISTS signal.
The solid upper line is the computed inverse average time 1/$<\tau>$
of the negative component. 

In conclusion, a NET model,
based on the assumption of local thermodynamic equilibrium
in an extended parameter space, 
accounts for the rich phenomenology
observed in the ISTS experiments. 
In particular, using new ISTS data
on supercooled glycerol in the temperature region where the
structural relaxation and the thermal diffusion process take
place on the same time scale, we have demonstrated that the model is
able to reproduce the experimental data using literature data from other experiments.
Further investigations on the possibility of assuming local thermodynamic
equilibrium in supercooled liquids are crucial in the development 
of a thermodynamic description of glass-forming liquids.
This work was supported by INFM, MURST and EC grant N. HPRI-CT1999-00111.


\begin{references}
\bibitem{note1}
Inverse average time $<\tau>^{-1}=\gamma \beta / \Gamma(1/\beta)$
is used as characteristic rate of the stretched exponential.
\bibitem{note3}
The number N should not be read as a number of well defined
physical variables, but rather as the discrete representation
of either a distribution of independent debye relaxing variables,
or a single strongly non exponential relaxing variable.
It  is important  to note that the specific studies and experiments  presented in this 
work do not allow to find whether the underlying relaxation 
process is homogeneous and  intrinsically  non exponential or it is heterogeneous, 
i.e. the superposition of different, spatially separated, genuine Debye processes (see \cite{richert} for a comprehensive
review on the subject).
\bibitem{notaLongTimeApprox} $\partial^2\rho/\partial t^2+(\eta q^2/\rho_0)\partial\rho/\partial t+q^2 p=0$.
After the acoustic transient the terms involving
derivatives of density are negligible.
\bibitem{noteN} Given the constraints in the model the number of input parameters
does not change with N.
We remark that, for N large enough ($>10$), the solution is practically independent of N.
\end{references}
\end{document}